\documentclass[conference]{IEEEtran}
\IEEEoverridecommandlockouts
\usepackage{cite}
\usepackage{amsmath,amssymb,amsfonts}
\usepackage{graphicx}
\usepackage{textcomp}
\usepackage{xcolor}
\usepackage{pifont}
\usepackage{balance}
\usepackage{xspace}
\usepackage{caption}
\usepackage{subcaption}
\usepackage{cite}
\usepackage{bm}
\usepackage{array}
\usepackage{setspace}
\usepackage[nameinlink]{cleveref}
\usepackage{mathtools}
\usepackage{amsthm}
\usepackage{multirow}
\usepackage{algorithm}
\usepackage{algpseudocode}
\usepackage{algcompatible}
\usepackage{tabularx}
\usepackage{color}
\usepackage{epstopdf}
\usepackage{endnotes}
\usepackage{framed}
\usepackage{colortbl}
\usepackage{cool} 
\usepackage{enumerate} 
\usepackage[bottom]{footmisc}
\usepackage{pifont}
\usepackage[super]{nth}
\usepackage[english]{babel}
\usepackage{multirow}
\usepackage{makecell}
\usepackage{soul}
\Crefname{figure}{Fig.}{Figs.}

\usepackage[subnum]{cases}

\newtheorem{assumption}{Assumption}

\usepackage{mathrsfs}
\usepackage[justification=centering]{caption}
\allowdisplaybreaks[4]

\newcolumntype{P}[1]{>{\centering\arraybackslash}p{#1}}

\DeclareUnicodeCharacter{2212}{-}

\usepackage{color}

\usepackage{titlesec}

\usepackage{mathrsfs}
\usepackage[justification=centering]{caption}
\allowdisplaybreaks[4]
\newcolumntype{P}[1]{>{\centering\arraybackslash}p{#1}}

\DeclareUnicodeCharacter{2212}{-}

\usepackage{color}

\begin{document}
\title{\LARGE Client Selection Approach in Support of Clustered Federated Learning over Wireless Edge Networks}
\author{\IEEEauthorblockN{Abdullatif Albaseer, Mohamed Abdallah, Ala Al-Fuqaha, and Aiman Erbad}
\IEEEauthorblockA{Division of Information and Computing Technology, College of Science and Engineering,
\\Hamad Bin Khalifa University, Doha, Qatar \\
\{amalbaseer, moabdallah, aalfuqaha, AErbad\}@hbku.edu.qa}
}
\maketitle

\begin{abstract}
Clustered Federated Multitask Learning (CFL) was introduced as an efficient scheme to obtain reliable specialized models when data is imbalanced and distributed in a non-i.i.d. (non-independent and identically distributed) fashion amongst clients.
While a similarity measure metric, like the cosine similarity,  can be used to endow groups of the client with a specialized model, this process can be arduous as the server should involve all clients in each of the federated learning rounds. 
Therefore, it is imperative that a subset of clients is selected periodically due to the limited bandwidth and latency constraints at the network edge. 
To this end, this paper proposes a new client selection algorithm that aims to accelerate the convergence rate for obtaining specialized machine learning models that achieve high test accuracies for all client groups. 
Specifically, we introduce a client selection approach that leverages the devices' heterogeneity to schedule the clients based on their round latency and exploits the bandwidth reuse for clients that consume more time to update the model. 
Then, the server performs model averaging and clusters the clients based on predefined thresholds. When a  specific cluster reaches a stationary point, the proposed algorithm uses a greedy scheduling algorithm for that group by selecting the clients with less latency to update the model.  
Extensive experiments show that the proposed approach lowers the training time and accelerates the convergence rate by up to $50\%$ while imbuing each client with a specialized model that is fit for its local data distribution.
\end{abstract}


\IEEEpeerreviewmaketitle

\section{INTRODUCTION}
\IEEEPARstart{I}{n} wireless edge networks, Internet of Things (IoT) devices produce a massive volume of data that can be used to understand the state of the system or predict its future states. 
Mobile Edge Computing can utilize the unprecedented capabilities of edge devices and access points (APs) to perform complex tasks and enable intelligent services \cite{park2019wireless}. 
\emph{Artificial Intelligence} (AI), and \emph{Machine Learning} (ML) specifically, have witnessed high-speed developments and started to introduce intelligent services that can revolutionize our lives.
Recently, several works have investigated the utilization of ML algorithms for future wireless edge applications, considering edge networks as an enabling technology for this vision\cite{lu2018mimo,luo2020power}. 

However, conventional ML techniques necessitate data to be offloaded and stored in a single central server which raises significant privacy concerns, increases the volume of data generated by edge devices, and the transmission delay of the limited communication resources when uploading data to the centralized server \cite{mao2017survey}. 

In response, Google proposed a promising decentralized machine learning technique termed Federated learning (FL) that addresses these problems and keeps the data where it is produced. 
The learning algorithm is run within edge devices, and only model parameters are shared with the server\cite{mcmahan2017communication} which initially broadcasts the latest version of the model for the update process. The server coordinates this process by aggregating and averaging all updates to form the global model. The server repeats these steps until the global model converges.
Note that we refer to FL as Federated Edge Learning (FEEL) throughout this paper as our work focuses on deploying FL in wireless edge networks.  

In fact, there are two major challenges, statistical heterogeneity and resource heterogeneity, when deploying FEEL.
For the statistical heterogeneity, the data across the network is massively distributed in non-i.i.d and imbalanced fashion\cite{mcmahan2017communication,zhao2018federated,9473806,albaseer2021fine}. 
As for the resource heterogeneity challenge, devices at the wireless edge are heterogeneous, and the bandwidth is limited; therefore, restricting the number of devices participating in a particular FEEL round. This heightens the need for efficient client selection strategies that contribute to a faster convergence rate, especially for large-scale edge networks.

To address the statistical challenges, a new approach called Clustered Federated Learning (CFL) has been proposed\cite{sattler2020clustered,sattler2020byzantine}. The multi-task clustering method utilizes the geometric characteristics of the FEEL loss surface to group the clients into clusters with federally trainable data distributions. 
It is worth noting that traditional FEEL assumes that only a single model can be trained on the consolidated data of all users. 
However, Sattler et al. \cite{sattler2020clustered} demonstrated that this assumption is frequently violated in practical FEEL applications. 
Concretely, it is shown that at every stationary point of the FEEL objective, the cosine similarity between the weight-updates of diverse clients can be used to infer the groups of clients having incongruent distributions. 
The main advantage of CFL, alongside multi-task FEEL \cite{smith2017federated} and Personalized Federated learning \cite{zhang2020personalized}, is that CFL does not require any modifications of the core FEEL communications protocol as clustering is only performed after the FEEL model converges to a stationary point. CFL can be seen as a post-processing method that can improve FEEL performance by providing clients with more specialized models. 
Recently, the studies in \cite{xie2020multi, ghosh2020efficient, briggs2020federated} followed up on the CFL-based framework and confirmed that CFL is a more reliable choice to gain high accuracy than the conventional FEEL.  

To address the resource heterogeneity, recent works~\cite{xu2020client, shi2020joint,yang2019scheduling,shi2020device,wadu2020federated} have attempted to address the problem of joint client scheduling and resource allocation considering non-i.i.d data distribution. For instance, the authors in \cite{xu2020client, shi2020joint,yang2019scheduling,shi2020device,wadu2020federated} proposed different algorithms to select the participants during FEEL rounds. 
The proposed selection algorithms choose the participants that either provide less updating time, have more powerful capabilities, or consume less energy, assuming that the number of data points is balanced for all devices. However, none of these works~\cite{xu2020client, shi2020joint,yang2019scheduling,shi2020device,wadu2020federated} consider the scheduling problem for the CFL approach, which initially needs to involve all active participating clients to perform local training updates and capture their local data distributions, making existing solutions inapplicable.

Spurred by the above observations, this work introduces a new client selection approach for CFL to minimize the training time, accelerate the convergence rate while tackling non-i.i.d and unbalanced data distribution problems. 
We propose an efficient client selection approach in which all clients have equal probability of participating in the training phase even if they have bad channels or low computational capabilities to obtain unbiased models. 
Specifically, we formulate an optimization problem and devise a client selection approach that achieves this objective.  
Experimental results verify that the proposed scheduling approach significantly minimizes the training time and accelerates the convergence rate.

The rest of the paper is structured as follows: we introduce the system, learning, computation, and communication models in Section~\ref{sysmodel}. 
The problem statement is formulated in Section ~\ref{sec:Problemformulation}. Section~\ref{Sec:proposed} presents the details of the proposed approach.
Experimental results are presented in Section~\ref{sec:experiment} where we discuss application scenarios. Finally, Section ~\ref{conclusion} concludes this work and provides directions for future extensions. 

\section{PRELIMINARIES}
\label{sysmodel}
\subsection{System Model}
\begin{figure}[t]
  \centering
  \includegraphics[width=0.9\linewidth]{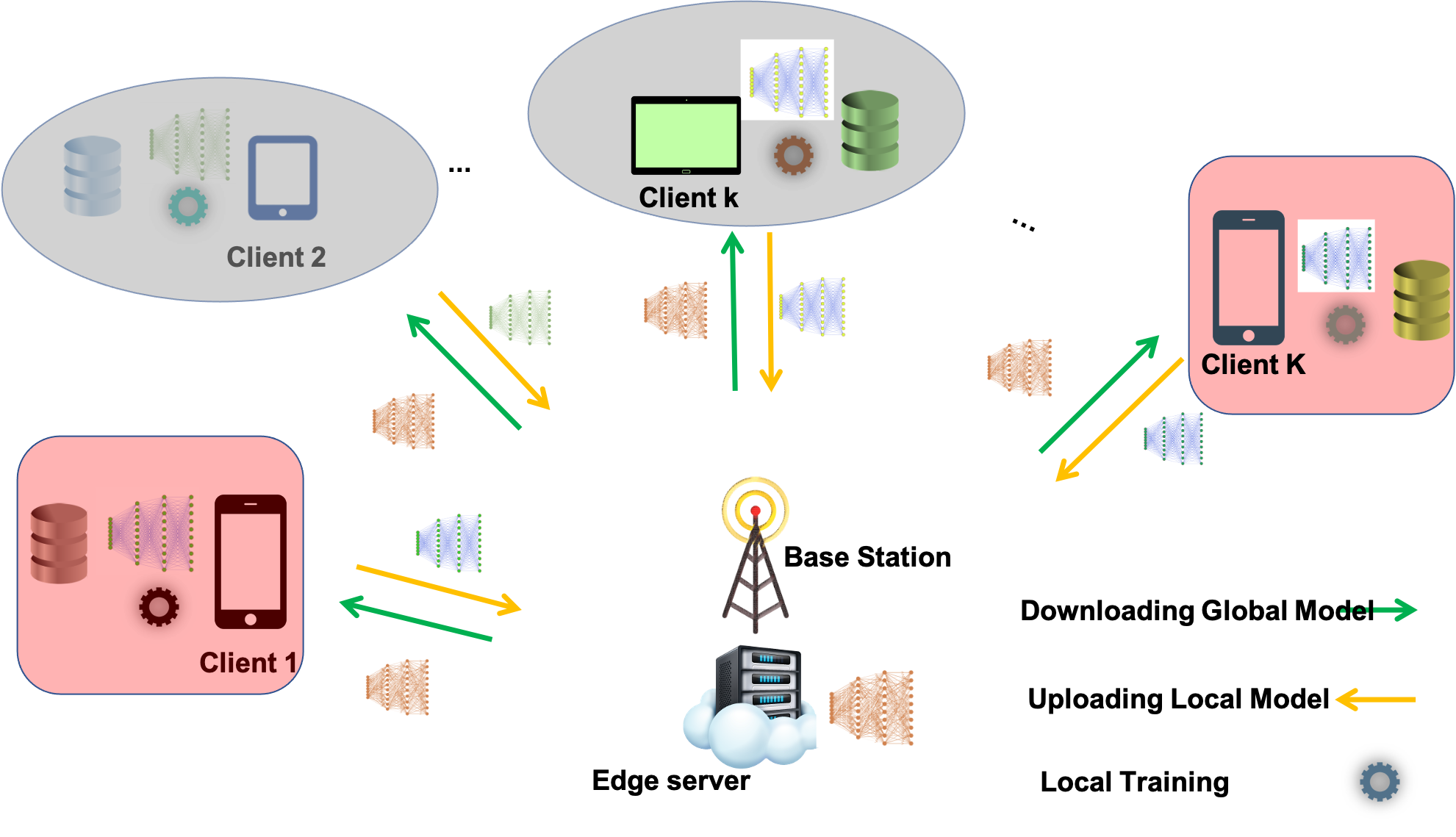} \setlength\belowcaptionskip{0cm}
  \caption{Clustered Multitask Federated Learning over wireless edge networks.}
  \label{WFEEL}
\end{figure}


As shown in \cref{WFEEL}, we consider a FEEL  across a set of $\mathcal{K}$ edge devices, $K = |\mathcal{K}|$, connected to one edge server which is equipped with a Base Station(BS). Each  $k \in \mathcal{K}$ device holds a local dataset $\mathcal{D}_k$, with number of samples denoted by $D_k = |\mathcal{D}_k|$. Thus, the total data samples of all clients is $D=\sum_{k=1}^{K}{D_k}$. Each local dataset $\mathcal{D}_k$ reprsnts the collection of data samples with input-output pairs as $\{\mathbf{x}_i^{(k)},y_i^{(k)}\}^{D_k}_{i=1}$, where $\mathbf{x}_i^{(k)} \in \mathbb{R}^d$ is an input vector with $d$ features, and $y_i^{(k)} \in \mathbb{R}$ is the associated class-labeled output $\mathbf{x}_i^{(k)}$. Due to different device activity, the data distribution of each individual represents its local data, not the population; thus the $\mathcal{D}_k$'s is non-i.i.d and imbalanced.
To train the model, each $k$-\textit{th} participant runs its local solver, such as stochastic gradient descent (SGD), locally to minimize the loss function for several local epochs denoted by  $E$.
For example, if a minibatch SGD is used, the total number of local updates is defined as: $\mathbf{n} = E~\frac{D_k}{b}$ where $b$ is a batch-size determining a subset of the training set required for one local update.

\subsection{FEEL model}

The edge server aims to learn a shared global model that can perform well across the network by leveraging the data of the $K$ connected devices. To perform this task, FEEL carries out the following steps every $r$\textit-th round: 1) The server initializes random model parameters ${w^{0}}$ at the beginning of the training process and selects a subset $\mathcal{S}^r$ of the clients to take part in the global model update by conducting local training, 2) The server broadcasts the latest global FEEL model parameters ${w^{r}}$
to all selected participants, 3) The selected participants in $\mathcal{S}^r$ upload their
updated models. Last, the edge server aggregates and fuses all updates to form a new version of the global model. Formally, the edge server coordinates to fit the global model parameters ${w^{r}} \in \mathbb{R}^d$ to capture the associated output $y_i$, by iteratively minimizing a local loss function $f_i({w^{r}})=\ell(\mathbf{x}_i^{(k)},y_i^{(k)}; {w^{r}})$ in the $r$-th communication round. The local loss function is defined as:
$
    F_k({w^{r}}):= \frac{1}{D_k}\sum_{i \in \mathcal{D}_k} 
    f_i({w^{r}}).
$
Accordingly, given datasets $D_1,..,D_K$ distributed over $K$ different clients, the goal in FEEL is to minimize the joint value of the loss function overall client's data $\mathcal{D}=\cup_k \mathcal{D}_k$ according to:

\begin{footnotesize}
\begin{align}
\label{eq:risk}
\min_w F(w, \mathcal{D}):=\frac{1}{D}\sum_{(x,y)\in \mathcal{D}}l(F_w(x),y)=\sum_{k=1}^K\frac{D_k}{D}\underbrace{F_k({w^{r}}, \mathcal{D}_k)}_{\text{data local sub-problem}}
\end{align}
\end{footnotesize}
\subsection{Computation and Communication Models}
Due to limited bandwidth in FEEL, the BS cannot accept all existing clients as the number of sub-channels is limited. Thus, in conventional FEEL, only a subset of these devices indicated as elected set $\mathcal{S}_r$ is selected every $r$-th round. Particularly, $\mathcal{S}_r=\{k \mid S_k^r=1, k=1,2,...,K\}$, where $S_k^r=1$ means that device $k$ is selected and added to $\mathcal{S}_r$, otherwise $S_k^r=0$.
The round latency of each device $k \in \mathcal{S}_r$ consists of two parts, i.e., communication latency and computation latency. In general, all devices have the same model structure initialized by the coordinating server, with fixed-size ${w^{r}}$ denoted by $\zeta$. Without the loss of generality, we assume that the orthogonal frequency division multiple access (OFDMA) scheme is employed and in every $r$-th round, each $k$-\textit{th} device is assigned an allocated bandwidth $\lambda_{k}^{r} B$ with one sub-channel where the system bandwidth is divided into $N$ sub-channels equals to the number of selected clients. Thus, the achievable data rate for each $k$ can be defined as: $ r_k^r=\lambda_k^r B\ln{\left( 1+\frac{P_{k}^r {h_{k}^{r}}^2}{N_0} \right)}, $
where $h_{k}^r$ denotes the channel gain between device $k$ and the BS, $P_{k}^r$ denotes the $k$-\textit{th} device's transmit power, and $N_0$ denotes spectral power density (i.e, Gaussian noise). Therefore, the transmit latency can be defined as: $
    T_k^{trans}= \frac{\zeta}{ r_k^r} \cdot, \label{trans_latency} 
$
On the other hand, the local computation latency of every device $k$ to train its local model can be expressed as:
$
  T_k^{cmp}= E\frac{\phi_k D_k}{f_k}
$
where $\phi_k$ denotes the number of required CPU cycles to process one sample;  $\phi_k D_k$ denotes the number of required CPU cycles  to process all samples in $r$-th round; and $f_k$ is the CPU frequency. Therfore, the total latency for every $k$-th device in the $r$-th round can be written:
\begin{equation}
     T_k^{total}= T_k^{trans} + T_k^{cmp}
\label{total_T} 
\end{equation}
In practice, the FEEL round latency, i.e., round deadline constraint, is determined by the latency of the slowest selected device $k \in \mathcal{S}_r$. Mathematically, we can define the round latency as:
$ T_{r} = \max\{T_k^{trans} + T_k^{cmp}\}$
\subsection{CFL Mechanism}
As opposed to the standard FEEL framework where all clients are treated equally to train only one global model, the goal of CFL is to provide every client with a model that optimally fits its local data distribution.  To consolidate the earlier exhibited problems with non-i.i.d and imbalanced data distributions, CFL~\cite{sattler2020clustered} aims to generalize the conventional FEEL assumption to a set of clients having the same data distribution: 
\begin{assumption}
\label{ass:2}
\textbf{("CFL")~\cite{sattler2020clustered}:} 
\textit{
There exists a partitioning $\mathcal{M}=\{c_1,..,c_m\}$, $\bigcup_{i=1}^M c_i = \{1,..,K\}$ of the client population, in such  a way that each subset of clients $c\in\mathcal{M}$ satisfies the standard FEEL assumption. 
}
\end{assumption}
\noindent where $\mathcal{M}$ is a set of clusters in which each cluster includes the clients having the same data distribution (i.e., congruent data distribution) and $M = |\mathcal{M}|$ is the number of clusters. 
In CFL, the FEEL system can be seen as an exemplary parameter tree. At the root node resides the standard FEEL model, obtained by converging to a stationary point $w^*$ of the FEEL objective. According to their cosine similarities, the client population is split up into two groups in the next layer, and each subgroup repeatedly converges to a stationary point $w^*_0$ and $w^*_1$, respectively. Branching lasts recursively till no stationary solution meets the partitioning criteria.
This is worth noting that the \emph{cosine similarity}, $ sim$, between the updates of any two clients $k$ and ${k'}$ is obtained by:

\begin{footnotesize}
\begin{align}
\begin{split}
 sim_{k,{k'}}:= sim(\nabla F_k(w^*), \nabla F_k'(w^*))&:=\frac{\langle\nabla F_k(w^*), \nabla F_{k'}(w^*)\rangle}{\|\nabla F_k(w^*)\|\|\nabla F_{k'}(w^*)\|}\\
\label{eq:cossim}
&=\begin{cases}
1 & \text{ if }I(k)=I(k')\\
-1 & \text{ if }I(k)\neq I(k')
\end{cases}
\end{split}
\end{align}
\end{footnotesize}

where $I(k)$ and $I(k')$ are the local data distribution of $k$ and $k'$, respectively.
Therefore, an accurate bi-partitioning is given by: 
$
c_1=\{k| sim_{k,0}=1\},~c_2=\{k| sim_{k,0}=-1\}.
$
According to \cite{sattler2020clustered}, the split should only perform if the following conditions are satisfied \cite{sattler2020clustered}:
\begin{align}
\label{eq:servernorm}
0 \leq \|\sum_{k=1,..,K}\frac{D_k}{D_m}\nabla_w F_k(w^*)\|<\varepsilon_1
\end{align}   
Equ. \eqref{eq:servernorm} holds that the solution is closer to a stationary point of the FEEL objective.  Conversely, the participating clients are away from a stationary point of their local loss if the following condition is satisfied:
\begin{align}
\label{eq:clientnorm}
\max_{i=k,..,K} \|\nabla_w F_k(w^*)\|>\varepsilon_2 > 0
\end{align}
where $\varepsilon_1$, and $\varepsilon_2$ are hyperparameters to control the clustering task.

\section{PROBLEM FORMULATION}
\label{sec:Problemformulation}
In CFL, it is assumed that either all clients are selected from the beginning of the training as in \cite{sattler2020clustered} or only a random subset is selected at each round as in \cite{FedGroupDuan}. The former is impossible due to the limited resources at the wireless network edge.   
This latter will not capture all incongruent data distribution among clients, inducing a problem of efficiently designing a client selection approach while satisfying the edge wireless network constraints and achieving desired performance. Specifically, to adequately theorize the classical FEEL setting, it is crucial to split up clients with incongruent data distributions at the earliest to accelerate the convergence rate and reduce the communication costs considering the limited bandwidth and the data and device heterogeneity. Hence, we formulate an optimization problem to find the optimal scheduling for a set of clients (i.e., $\mathcal{S}_{r}$) that accelerates the convergence rate and minimizes the training time for all trained models (i.e., $M$ groups models and one FEEL global model).
Concretely, we use $R$ to indicate the whole rounds during the training time budget $T^{total}$. We aim to find the optimal set of $M$ models ${{w_m}}, m=1, 2, \dots, M$ within $T^{total}$ that minimize the global empirical risk function for each cluster as follows: 
\begin{equation}
\setlength\abovedisplayskip{3pt}
\setlength\belowdisplayskip{3pt}
    {{w_m}} \triangleq \underset{{w}\in \{ {w}_r^{{\mathcal{S}_{r}}}:r=1,2,\dots,R\}}{\text{arg\,min}} F({w_m}).
    \label{def-tildew}
\end{equation}
To ease the presentation, we use $[R]$  to denote $\{1, 2, \dots, R\}$. The optimization problem for all clusters can be written as follows:
\begin{align}
\setlength\abovedisplayskip{3pt}
\setlength\belowdisplayskip{3pt}
    \underset{R, {\mathcal{S}_{[R]}}, {T}_{[R]}}{\text{min}} \quad & T^{total} \tag{P1}\label{P1}\\
    \text{s.t.} \qquad \;\; \quad & F({{w_m}}) - F({{w^*_m}}) \le \epsilon,  \forall m \in \mathcal{M},  \tag{P1.0}\label{P10} \\ 
    & \sum_{r=1}^R T_{r}(\mathcal{S}_r) \leq T,  \tag{P1.1}\label{P11}\\
    & T_k^{total}\leq T_r(\mathcal{S}_r),  \tag{P1.2}\label{P12} \\
    & \mathcal{S}_r \subset \mathcal{K}, \forall r \in [R], \forall m \in \mathcal{M} \tag{P1.3}\label{P13} \\
    & |\mathcal{S}^m_r| \le N, \forall r \in [R], \tag{P1.4}\label{P14}
\end{align}
where ${\mathcal{S}_{[R]}} = [\mathcal{S}_1, \mathcal{S}_2, \dots, \mathcal{S}_R]$ is the selected scheduling set for all rounds and ${T}_{[R]} = [T_1, T_2, \dots, T_R]$ is the maximum latency for each round (i.e., deadline constraint). 
 Constraint~\eqref{P10} guarantees that the convergence of each group model is satisfied. Constraint \eqref{P11} is set to assure that the total training time of all rounds doesn't exceed the time budget. At the same time, constraint~\eqref{P12} ensures that the computation and communication time for every participating client is aligned with the deadline constraint. 
Constraint~\eqref{P14} stipulates that the allocated bandwidth for all selected clients can not exceed the system bandwidth. To find the solution of \ref{P1}, we first need to know the effects of $R$ and ${\mathcal{S}}_{[R]}$ on the loss function of each cluster model, i.e., $F({{w_m}})$. 
It is worth noting that finding an exact expression of $F({{w_m}})$ considering $R$ and $\mathcal{S}_{[R]}$ is impossible. Thus, we tend to solve $F({{w_m}})$ iteratively in terms of $R$ and $\mathcal{S}_{[R]}$. In addition, due to the variation of computation delay, $T_k^{cmp}$ and wireless channel states $h_{k}^r$ through the rounds $R$, attaining the optimal client selection approach is challenging and could be non-stationary. In reality, the FEEL process is iterative in nature and depends on the scheduled clients in the preceding rounds, making the bounding of $F({{w_m}})$ also very hard. To this end, we solve P1 in the next sections as follows.
First, we extend the algorithm \cite{sattler2020clustered} considering the computation and communication resources. Then, we propose a fair client scheduling and resource allocation approach to solve the issues that emanate from having to schedule all clients in every round to avoid obtaining biased models.
\setlength{\textfloatsep}{1pt}
\section{PROPOSED SCHEDULING ALGORITHM}
\label{Sec:proposed}
In our proposed scheduling algorithm, all active and available clients during the training phase have equal priority to participate in the training. 
Thus, the server will collect preliminary information from these devices, such as data size, local CPU speed, and channel state; then, the server evaluates the expected latency of all clients to schedule the model uploading based on their finishing time. 
To aggregate the updates in an efficient manner, the server sorts the clients based on their expected latency in ascending order. At this stage, 
The number of aggregations that the server needs to collect all updates in one round can be defined as:
\begin{equation}
\label{eq:ng}
   ng = \frac{|{\mathcal{S}_{[r]}}|}{N}
\end{equation}
As some participating clients complete updating and uploading tasks, other clients can perform their updates while the former clients are in the upload phase. Therefore, the total time budget is efficiently utilized as the server can handle many updates subsequently. The aggregation set can be defined as:

\begin{footnotesize}
\begin{align}
\label{eq:gset}
     \mathcal{G}  = &\{\{1+(N+(j-1)), \dots, N+(N+(j-1))\}, j=1, 2, \dots, ng\}
\end{align}
\end{footnotesize}

\begin{algorithm}[]
\footnotesize
\caption{Client Selection Strategy for CFL}
\label{alg:proposedscheduling}
\begin{algorithmic}[1]
\STATEx \textbf{Input}: $K$ (number of clients), initial parameters $w_0$, controlling parameters $\varepsilon_1, \varepsilon_2 > 0$, number of local epochs $E$, batch size $b$
\STATEx \textbf{Output}: $M$ reliable models, one FEEL global model
\STATEx \textbf{Initialization:} set initial clusters $\mathcal{M}=\{ \{ 1,..,K\} \}$,  initial models $w_k \leftarrow 0$  $\forall k$, set $r=1$
\WHILE {Not converge}
\begin{center}
\textbf{\textit{// Server Task Prepossessing}}
\end{center}
\STATE Server \textbf{collects} prior information from all active and available clients
\IF{$M > 1$ }
\STATE Server \textbf{selects} the best clients for all $M-1$ clusters while \textbf{selects} all client in last cluster, $M$, from the cluster set $\mathcal{M}$
\ELSE 
\STATE Server~\textbf{selects} all active and available clients 
\ENDIF
\STATE Server \textbf{estimates} the latency of all clients and sort them in ascending order to schedule uploading the updates 
\STATE Server \textbf{finds} the aggregation set using \eqref{eq:ng} and \eqref{eq:gset}
\STATE Server \textbf{broadcasts} $w_{r-1}$ to all clients
\STATEx
\begin{center}
\textbf{\textit{// Participating Clients Task in parallel}}
\end{center}
\FOR {$k$ = 1 to $|{\mathcal{S}_{[r]}}|$}
\STATE \textbf{Receive} $w_{r-1}$ from the server
\STATE \textbf{Perform} local updates  $ w_k = \boldsymbol  w^{r-1} - \eta \sum_{i=1}^{\mathbf{n}} \nabla F_k^{r}(\boldsymbol  w_k(i))$ 
\ENDFOR
\STATE Server \textbf{aggregates} all updates as ordered in $\mathcal{G}$ (i.e., Equ \eqref{eq:gset})
\STATEx
\begin{center}
\textbf{\textit{// Server Task Post-processing}}
\end{center}
\STATE Server \textbf{sets} $\mathcal{M}_{tmp} \leftarrow \mathcal{M}$
\STATE Server \textbf{computes} $F({w^{r}}) =\sum_{k=1}^{K}\frac{D_k}{D}{ F_k({w^{r}})}$ and ${w^{r}} =\sum_{k=1}^{K}\frac{D_k}{D}{ {w^{r}}}$

\FOR {$c$ $\in$ $\mathcal{M}$}
\STATE \textbullet~ Server \textbf{finds} $\Delta w_m \leftarrow \frac{1}{|c|}\sum_{k\in c}\Delta w_k$ 
\IF {$\|\Delta w_c\|<\varepsilon_1$  \textbf{and} $\max_{k\in c}\|\Delta w_k\|>\varepsilon_2$}
\STATE \textbullet~ $ sim_{k,{k'}}\leftarrow\frac{\langle \Delta w_k,\Delta w_{k'}\rangle}{\|\Delta w_k\|\|\Delta w_{k'}\|}$
\STATE \textbullet~ $c_1,c_2\leftarrow \arg\min_{c_1\cup c_2 = c}(\max_{k\in c_1, k'\in c_2}  sim_{k,k'})$
\STATE \textbullet~ $ sim_{cross}^{max} \leftarrow \max_{k\in c_1, {k'}\in c_2} sim_{k,k'}$
\STATE \textbullet~ $\gamma_k:=\frac{\|\nabla F_{I(k)}(w^*)-\nabla F_k(w^*)\|}{\|\nabla F_{I(k)}(w^*)\|}$
\IF{$max(\gamma_{k}) < \sqrt{\frac{1- sim_{cross}^{max}}{2}}$}
\STATE \textbullet~ $\mathcal{M}_{tmp}\leftarrow(\mathcal{M}_{tmp}\setminus c) \cup c_1\cup c_2$
\ENDIF
\ENDIF
\ENDFOR

\STATE \textbullet~ $\mathcal{M} \leftarrow \mathcal{M}_{tmp}$
\STATE $r = r+1$
\ENDWHILE
\STATE \textbf{Server \textit{Return}} $M$ reliable models, one FEEL global model
\end{algorithmic}
\end{algorithm}
It is worth noting that the server is assumed to have far greater computational resources
than the participating clients; thus, the cost related to this task will typically be negligible. 
Our scheduling problem has a trade-off between the number of clients selected in a specific round and the total number of rounds handled in a particular duration, which can be controlled by $T_{r}$. If we initiated long $T_{r}$, the number of participating clients in $\mathcal{S}_{[r]}$ can increase, which is desired for CFL at the round and can expedite the splitting and improve the performance of the group models. Moreover, longer $T_{r}$ decreases the number of rounds performed in a particular duration, which can catch the incongruent degree of data distribution. Thus, we should carefully select $T_{r}$.
At each round, the server uses \eqref{eq:cossim}, \eqref{eq:servernorm}, and \eqref{eq:clientnorm} to partition and cluster the clients. In such a case, we have the the similarities between participating clients $k$ and ${k'}$ within a cluster which can be bounded as:  
\begin{align}
\label{eq:within}
 sim_{within}^{min} = \min_{\underset{I(k)=I({k'})}{k,{k'}}} sim(\nabla_w r_k(w^*), \nabla_w r_{k'}(w^*))
\end{align}
While the similarities between participating clients $k$ and ${k'}$ across different clusters (i.e., different data distributions) can be bounded as:
\begin{align}
\label{eq:cross}
 sim_{cross}^{max}=\max_{k\in c_1^*,{k'}\in c_2^*} sim(\nabla_w r_k(w^*), \nabla_w r_{k'}(w^*))
\end{align} 
We should emphasize that \eqref{eq:within} and \eqref{eq:cross} are used to estimate the separation gap as follow:
\begin{align}
g( sim):= sim_{intra}^{min}- sim_{cross}^{max} 
\end{align} 
The steps of the proposed client selection approach are listed in Algorithm \ref{alg:proposedscheduling}. In Alg.~\ref{alg:proposedscheduling}, we can see that once a particular cluster reaches the stationary point of FEEL objective (lines 2-6), the server uses greedy scheduling for such clusters to choose clients providing less training time to perform further updates. This stems from that the clustered clients, in this case, have congruent distribution.

\section{NUMERICAL EXPERIMENTS}
In this section, we present the experimental setup, the simulated scenarios, and the numerical results.
\label{sec:experiment}
\subsection{Experimental Settings}
  We consider a total bandwidth of $B=10$ MHz, and the bandwidth of each sub-channel is $1$ MHz. We model a random wireless channel gain $h_{k}^r$ for each participant with a path loss ($\mu = g_0(\frac{d_0}{d})^4$) where $g_0 = −35 $ dB and the reference distance $d_0 = 2$ m. The distances between edge devices and the edge server are distributed uniformly between $20$ and $100$ m. Also, Additive White Gaussian Noise (AWGN) power is set to $N_0=10^{-6}$ watt. The transmit power $P_{k}^r$ is randomly distributed between $p_{min} = -10$ dBm and $p_{max} = 20$ dBm. The CPU frequency $f_k$ is randomly distributed between $1$ GHz and $9$ GHz, and the number of cycles $\phi$ required to process one sample is set to $20$. We use a federated dataset called, \textbf{FEMNIST} \cite{caldas2018leaf}, for classifications tasks. FEMNIST is used for the classification task of both letters and digits (A-Z, a-z, and 0-9), and it has 244,154 images for training and 61500 for testing. FEMNIST dataset \cite{caldas2018leaf} was specifically designed to model the challenges of realistic FEEL problems. We use this dataset under non-i.i.d data distribution, where the dataset is first split into 62 partitions, and then each user is assigned batches of two classes only. We use the convolutional neural networks (CNN) classifier to train the models.

\subsection{Experiments and Numerical Results}
To evaluate the performance of the proposed scheduling algorithm, we compared the results of the proposed algorithm against the random scheduling CFL algorithm (baseline algorithm as in \cite{sattler2020byzantine,FedGroupDuan}). To ensure a fair comparison, we use a similar setting in terms of several local updates, learning rate, model structure, data distribution (i.e., non-i.i.d and imbalanced data distribution) for both scheduling algorithms. We set $R$, the communication rounds, to 200 and the number of epochs $E$ to 10 for all clients. We run the experiments to training different models using both algorithms, and we train the models over 100 users and test the resulting models using the test data among $15$ clients.  

\begin{figure}[!t]
\centering
	     \begin{subfigure}[b]{0.5\textwidth}
         \centering
         \includegraphics[height= 3.5cm,width=\textwidth]{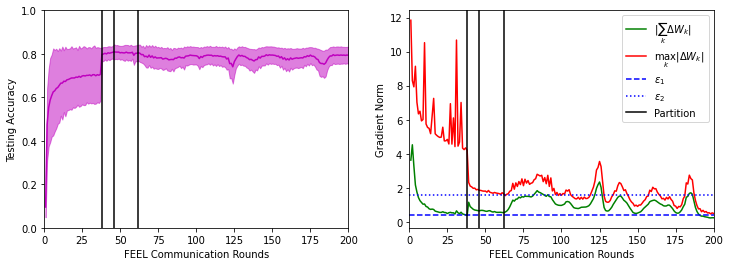}
         \caption{Proposed scheduling algorithm.}
         \label{F:a}
     \end{subfigure}
     \hfill
	     \begin{subfigure}[b]{0.5\textwidth}
         \centering
         \includegraphics[height= 3.5cm,width=\textwidth]{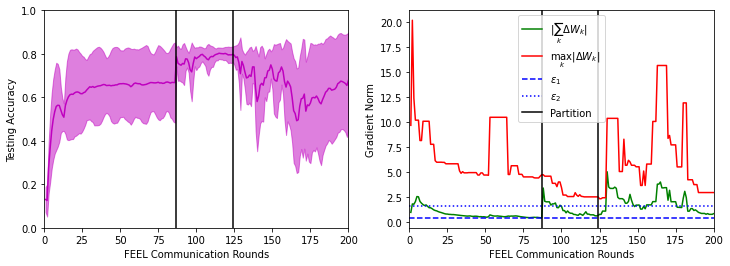}
         \caption{Baseline scheduling algorithm.}
         \label{F:b}
     \end{subfigure}
     \hfill     
	\caption{Average accuracy of the resulting
model(s) vs communication rounds using our proposed scheduling algorithm, top: (a: left), and using the baseline scheduling algorithm, Bottom: (b:left). In (a: right) the Gradient norm of global and local risk functions vs communication rounds using the proposed algorithms and (b:right) using the baseline algorithm. Results are collected and averaged over five trials.}
\label{F:acc}
\end{figure}

\Cref{F:acc} shows the performance of the proposed approach, and the baseline algorithms in terms of accuracy bounded with the confidence interval at every round (\Cref{F:a,F:b}, left) and the gradient norms to exhibit the convergence rate (\Cref{F:a,F:b}, right). One can see that the proposed algorithm gains much better accuracy with a faster convergence rate as the splitting started at round 37 while the split using the baseline algorithm started at round 83, accelerating the convergence rate by more than $50\%$. It is observed that, in the proposed algorithm, the partitioning is performed again for all clients with incongruent distributions at rounds 45 and 63 till all clusters were split. From \Cref{F:a}, right, we can note that all trained models reached the stationary point at round 190, indicating that clustering is finalized. At the same time, From \Cref{F:b}, right, the baseline algorithm, we can clearly observe that the FEEL still needs more rounds to converge to the stationary point at round 200. This stems from the behavior of random client selection, which might either select clients that have been clustered in former rounds or select the clients that exceed the deadline.

\setlength{\textfloatsep}{5pt}

\begin{table*}[htbp]
\caption{TESTING ACCURACY OF ALL RESULTING MODELS: A FEEL BASE MODEL AS WELL AS MORE SPECIALIZED MODELS FOR THE DIFFERENT CLUSTERS (THE PROPOSED SCHEDULING ALGORITHM AND THE BASELINE ALGORITHM), C*=CLIENT.}
\label{basetable}
 \begin{subtable}{\linewidth}
 \footnotesize
 \centering
 \caption{Proposed scheduling algorithm.}
   \label{table_proposedsetmodel}
\begin{tabular}{lllllllllllllllllllllll}
 \hline
  & C 1 & C 2 & C 3 & C 4 & C 5 & C 6 & C 7 & C 8 & C 9 & C 10 & C 11 & C 12 & C 13 & C 14 & C 15 \\ \hline
{\textbf{\textit{FEEL Model}}} & 46    & 37.9   & 45.2   & 44.8   & 37.3   & 75.5    & 80.7   & 83.2   & 79    & 79    & 77.3   & 76.6    & 77.9   & 76.4    & 77.5  &       \\ \hline
{\textbf{\textit{Model 1}}}  & 0        & 0        & 0        & 0        & 0        & \cellcolor{green!40}\textbf{77.1}   & 83.3   & \cellcolor{green!40}\textbf{86}   & \cellcolor{green!40}\textbf{81.6}   & 82    & \cellcolor{green!40}\textbf{81.8}    & \cellcolor{green!40}\textbf{81.5}    & 82.5    & 79.6     & 80.9    \\ \hline
{\textbf{\textit{Model 2}}}  & 0        & 0        & \cellcolor{green!40}\textbf{81.6}   & \cellcolor{green!40}\textbf{76}   & 77.6   & 0        & \cellcolor{green!40}\textbf{83.8}   & 85.5   & 0        & \cellcolor{green!40}\textbf{82.7}    & 80.1    & 0         & \cellcolor{green!40}\textbf{84.5}   & \cellcolor{green!40}\textbf{82}   & \cellcolor{green!40}\textbf{81.8}           \\ \hline
{\textbf{\textit{Model 3}}}  & \cellcolor{green!40}\textbf{83.3}   & 0        & 77   & 67   & \cellcolor{green!40}\textbf{77.9}   & 76.8   & 0        & 0        & 0        & \cellcolor{green!40}\textbf{82.7}    & 0         & 0         & 0         & 0         & 0                 \\ \hline
{\textbf{\textit{Model 4}}}  & 0        & 74.5   & 0        & 0        & 0        & 0        & 82.5   & 83   & 75.2   & 0         & 0         & 0         & 0         & 0         & 0                 \\ \hline
{\textbf{\textit{Model 5}}}  & \cellcolor{green!40}\textbf{83.3}  & \cellcolor{green!40}\textbf{78.6}   & 0        & 0        & 0        & 76.8   & 0        & 0        & 75.2   & 0         & 0         & 76.8    & 0         & 0         & 0               \\ \hline
{\textbf{\textit{Model 6}}}  & \cellcolor{green!40}\textbf{83.3}   & 74.5   & 76.9   & 67.2   & \cellcolor{green!40}\textbf{77.9}   & 0        & 0        & 0        & 0        & 0         & 0         & 0         & 0         & 0         & 0          \\ \hline 
\textit{\textbf{Max Acc}}  & \textbf{83.3}              & \textbf{78.6}              & \textbf{81.6}              & \textbf{76}              & \textbf{77.9}              & \textbf{77.1}                         & \textbf{83.8}                          & \textbf{86}                          & \textbf{81.6}                 &\textbf{82.7}           &\textbf{81.8}  & \textbf{81.5}                            & \textbf{84.5}                             & \textbf{82}                             & \textbf{81.8}                    \\ \hline 
\end{tabular}
\end{subtable} 

 \begin{subtable}{\linewidth}
 \footnotesize
 \centering
  \caption{Baseline scheduling algorithm.}
  \label{table_baselinesetmodel}
\begin{tabular}{lllllllllllllllllllll}
\hline
                  & \textit{\textbf{C 1}} & \textit{\textbf{C 2}} & \textit{\textbf{C 3}} & \textit{\textbf{C 4}} & \textit{\textbf{C 5}} & \textit{\textbf{C 6}} & \textit{\textbf{C 7}} & \textit{\textbf{C 8}} & \textit{\textbf{C 9}} & \textit{\textbf{C 10}} & \textit{\textbf{C 11}} & \textit{\textbf{C 12}} & \textit{\textbf{C 13}} & \textit{\textbf{C 14}} & \textit{\textbf{C 15}} \\ \hline 
\textit{\textbf{FEEL Model}} & 50.7                       & 41.6                       & 49.2                       & 46.4                       & 41.5                       & 73.1                       & 80.1                       & 82.1                       & 76.5                       & 77.8                        & 77                          & 77.3                        & 76.7                        & 76.1                        & 77.7                         \\ \hline 
\textit{\textbf{Model 1}}  & 0                          & 0                          & 0                          & 0                          & 0                          & 78.5                       & \cellcolor{green!40}\textbf{84.5}              & \cellcolor{green!40}\textbf{86.5}              & \cellcolor{green!40}\textbf{82.1}              & \cellcolor{green!40}\textbf{81.4}               & 81.5                        & \cellcolor{green!40}\textbf{79.9}               & \cellcolor{green!40}\textbf{81}                 & \cellcolor{green!40}\textbf{79.9}               & \cellcolor{green!40}\textbf{81.1}              \\ \hline 
\textit{\textbf{Model 2}}  & 0                          & 0                          & 0                          & \cellcolor{red!40}\textbf{52.4}              & \cellcolor{red!40}\textbf{56.5}              & 0                          & 78.3                       & 75.8                       & 77.9                       & 72.7                        & 0                           & 0                           & 78.7                        & 78.2                        & 67.4                        \\ \hline 
\textit{\textbf{Model 3}}  & \cellcolor{red!40}\textbf{59.7}              & \cellcolor{red!40}\textbf{61.1}              & \cellcolor{red!40}\textbf{62.5}              & 0                          & 0                          & \cellcolor{green!40}\textbf{82}                & 0                          & 0                          & 0                          & 0                           & \cellcolor{green!40}\textbf{82.8}               & 67.9                        & 0                           & 0                           & 0                           \\ \hline 
\textit{\textbf{Model 4}}  & \cellcolor{red!40}\textbf{59.7}              & \cellcolor{red!40}\textbf{61.1}              & \cellcolor{red!40}\textbf{62.5}              & \cellcolor{red!40}\textbf{52.4}              & \cellcolor{red!40}\textbf{56.5}              & 0                          & 0                          & 0                          & 0                          & 0                           & 0                           & 0                           & 0                           & 0                           & 0                          \\ \hline 
\textit{\textbf{Max Acc}}  & \textbf{59.7}              & \textbf{61.1}              & \textbf{62.5}              & \textbf{52.4}              & \textbf{56.5}              & \textbf{82}                         & \textbf{84.5}                          & \textbf{86.5}                          & \textbf{82.1}                 &\textbf{81.4}           &\textbf{82.8}  & \textbf{79.9}                            & \textbf{81}                             & \textbf{79.9}                             & \textbf{81.1}                    \\ \hline 
\end{tabular}
  \end{subtable} 
\end{table*}
From the presented results in Table \ref{basetable}, one can observe that the training process using both scheduling algorithms has resulted in multiple models for every client: A FEEL base model as well as more specialized models for the different clusters. Note that the proposed algorithm produces three more specialized models. From table \ref{table_proposedsetmodel}, one can notice that, in the results of the proposed algorithm, all clients attained satisfying accuracy (highlighted in green), and the performance gap 
is about $10\%$ only. In contrast, in table \ref{table_baselinesetmodel}, about $\frac{1}{3}$ of the tested clients (clients 1, 2, 3, 4, and 5) attained bad accuracy (highlighted in red). The performance gap reaches up to $30.4\%$, indicating that some models are still biased to those frequently selected clients. 
Overall, our results demonstrate that the proposed scheduling algorithm outperforms the baseline scheduling algorithm method in terms of clustering quality not only with more specialized models but also on accelerating the convergence rate and reducing the costs. 
\section{CONCLUSION}
\label{conclusion}
This work introduces a novel client selection approach for CFL to minimize the training time and accelerate the convergence rate. We first formulate an optimization problem to find the optimal client selection approach that reduces the training time and accelerates the convergence rate considering non-i.i.d and imbalanced data distribution, device heterogeneity, and limited bandwidth. The proposed client selection approach is fair to all clients, i.e., all clients have an equal probability of participating in the training phase to obtain unbiased models even if they have bad channels or few data points. We carry out extensive simulation experiments using a realistic federated dataset. The results verify that the proposed solution effectively accelerates the convergence rate by up to $50\%$ and minimizes the training time while attaining a more satisfying performance for each client with an optimal fit model for its local data distribution. Finding the optimal threshold values for splitting the clusters will be considered in future studies.
\section*{ACKNOWLEDGEMENT}
This publication was made possible by NPRP-Standard (NPRP-S) Thirteen (13th) Cycle grant \# NPRP13S-0201-200219 from the Qatar National Research Fund (a member of Qatar Foundation). The findings herein reflect the work and are solely the responsibility of the authors.
\bibliographystyle{IEEEtranTIE}
\bibliography{Ref}
\balance

\end{document}